\documentclass[aps,prd,twocolumn,showpacs,superscriptaddress,showpacs]{revtex4}  % for review and submission 
\usepackage{epsfig}
\usepackage{graphicx} 
\usepackage{dcolumn}  
\usepackage{bm}       
\usepackage{amssymb}  

\def\mevc {\ifmmode {\rm MeV}/c \else MeV$/c$\fi}
\def\mevcc {\ifmmode {\rm MeV}/c^2 \else MeV$/c^2$\fi}
\def\gevc {\ifmmode {\rm GeV}/c \else GeV$/c$\fi}
\def\gevcc {\ifmmode {\rm GeV}/c^2 \else GeV$/c^2$\fi}
\def\tevcc {\ifmmode {\rm TeV}/c^2 \else TeV$/c^2$\fi}

\def\vtd  {\ifmmode |V_{td}| \else $|V_{td}|$\fi}
\def\vtb  {\ifmmode |V_{tb}| \else $|V_{tb}|$\fi}
\def\vts  {\ifmmode |V_{ts}| \else $|V_{ts}|$\fi}
\def\vcb  {\ifmmode |V_{cb}| \else $|V_{cb}|$\fi}

\newcommand{\xs} {\ifmmode x_{\mbox{\sl s}}
                       \else $x_{\mbox{\sl s}}$\fi}
\newcommand{\xd} {\ifmmode x_d \else $x_d$\fi}
\newcommand{\lxy} {\ifmmode L_{\rm xy} \else $L_{\rm xy}$\fi}
\newcommand{\dgam} {\ifmmode \Delta\Gamma \else $\Delta\Gamma$\fi}
\newcommand{\dm} {\ifmmode \Delta m \else $\Delta m$\fi}
\newcommand{\ctau} {\ifmmode c\tau \else $c\tau$\fi}
 
\newcommand{\et}{{ E}_{ T}}

\newcommand{\abseta}{|\eta| \leq}

\newcommand{\met}{\mbox{$\protect \raisebox{.3ex}{$\not$}\et$}}

\newcommand{\ppbar}{ { p} \bar{{ p}} }

\newcommand{\ttbar}{t\bar{t}}

\newcommand{\mtop}{M_{t}}

\newcommand{\afb}{A_\mathrm{FB}}
\newcommand{\ac}{A_\mathrm{C}}
\newcommand{\afbl}{A_\mathrm{FB}^{\ppbar}}
\newcommand{\afbt}{A_\mathrm{FB}^{\ttbar}}

\newcommand{\ifb}{ {\rm fb}^{-1} }

\begin{document}

%\pagewiselinenumbers
% the following line is for submission 
\hspace{4.0in} \mbox{Fermilab-Pub-08-171-E, IEKP-KA/2008-21}
\bibliographystyle{apsrev}

%\leftline{Version 0.1 as of \today}

\title{Forward-Backward Asymmetry in Top Quark Production\\   
in $\ppbar$ Collisions at $\sqrt{s}=1.96$ TeV\\ }

\affiliation{Institute of Physics, Academia Sinica, Taipei, Taiwan 11529, Republic of China} 
\affiliation{Argonne National Laboratory, Argonne, Illinois 60439} 
\affiliation{University of Athens, 157 71 Athens, Greece} 
\affiliation{Institut de Fisica d'Altes Energies, Universitat Autonoma de Barcelona, E-08193, Bellaterra (Barcelona), Spain} 
\affiliation{Baylor University, Waco, Texas  76798} 
\affiliation{Istituto Nazionale di Fisica Nucleare Bologna, $^t$University of Bologna, I-40127 Bologna, Italy} 
\affiliation{Brandeis University, Waltham, Massachusetts 02254} 
\affiliation{University of California, Davis, Davis, California  95616} 
\affiliation{University of California, Los Angeles, Los Angeles, California  90024} 
\affiliation{University of California, San Diego, La Jolla, California  92093} 
\affiliation{University of California, Santa Barbara, Santa Barbara, California 93106} 
\affiliation{Instituto de Fisica de Cantabria, CSIC-University of Cantabria, 39005 Santander, Spain} 
\affiliation{Carnegie Mellon University, Pittsburgh, PA  15213} 
\affiliation{Enrico Fermi Institute, University of Chicago, Chicago, Illinois 60637} 
\affiliation{Comenius University, 842 48 Bratislava, Slovakia; Institute of Experimental Physics, 040 01 Kosice, Slovakia} 
\affiliation{Joint Institute for Nuclear Research, RU-141980 Dubna, Russia} 
\affiliation{Duke University, Durham, North Carolina  27708} 
\affiliation{Fermi National Accelerator Laboratory, Batavia, Illinois 60510} 
\affiliation{University of Florida, Gainesville, Florida  32611} 
\affiliation{Laboratori Nazionali di Frascati, Istituto Nazionale di Fisica Nucleare, I-00044 Frascati, Italy} 
\affiliation{University of Geneva, CH-1211 Geneva 4, Switzerland} 
\affiliation{Glasgow University, Glasgow G12 8QQ, United Kingdom} 
\affiliation{Harvard University, Cambridge, Massachusetts 02138} 
\affiliation{Division of High Energy Physics, Department of Physics, University of Helsinki and Helsinki Institute of Physics, FIN-00014, Helsinki, Finland} 
\affiliation{University of Illinois, Urbana, Illinois 61801} 
\affiliation{The Johns Hopkins University, Baltimore, Maryland 21218} 
\affiliation{Institut f\"{u}r Experimentelle Kernphysik, Universit\"{a}t Karlsruhe, 76128 Karlsruhe, Germany} 
\affiliation{Center for High Energy Physics: Kyungpook National University, Daegu 702-701, Korea; Seoul National University, Seoul 151-742, Korea; Sungkyunkwan University, Suwon 440-746, Korea; Korea Institute of Science and Technology Information, Daejeon, 305-806, Korea; Chonnam National University, Gwangju, 500-757, Korea} 
\affiliation{Ernest Orlando Lawrence Berkeley National Laboratory, Berkeley, California 94720} 
\affiliation{University of Liverpool, Liverpool L69 7ZE, United Kingdom} 
\affiliation{University College London, London WC1E 6BT, United Kingdom} 
\affiliation{Centro de Investigaciones Energeticas Medioambientales y Tecnologicas, E-28040 Madrid, Spain} 
\affiliation{Massachusetts Institute of Technology, Cambridge, Massachusetts  02139} 
\affiliation{Institute of Particle Physics: McGill University, Montr\'{e}al, Canada H3A~2T8; and University of Toronto, Toronto, Canada M5S~1A7} 
\affiliation{University of Michigan, Ann Arbor, Michigan 48109} 
\affiliation{Michigan State University, East Lansing, Michigan  48824}
\affiliation{Institution for Theoretical and Experimental Physics, ITEP, Moscow 117259, Russia} 
\affiliation{University of New Mexico, Albuquerque, New Mexico 87131} 
\affiliation{Northwestern University, Evanston, Illinois  60208} 
\affiliation{The Ohio State University, Columbus, Ohio  43210} 
\affiliation{Okayama University, Okayama 700-8530, Japan} 
\affiliation{Osaka City University, Osaka 588, Japan} 
\affiliation{University of Oxford, Oxford OX1 3RH, United Kingdom} 
\affiliation{Istituto Nazionale di Fisica Nucleare, Sezione di Padova-Trento, $^u$University of Padova, I-35131 Padova, Italy} 
\affiliation{LPNHE, Universite Pierre et Marie Curie/IN2P3-CNRS, UMR7585, Paris, F-75252 France} 
\affiliation{University of Pennsylvania, Philadelphia, Pennsylvania 19104} 
\affiliation{Istituto Nazionale di Fisica Nucleare Pisa, $^q$University of Pisa, $^r$University of Siena and $^s$Scuola Normale Superiore, I-56127 Pisa, Italy} 
\affiliation{University of Pittsburgh, Pittsburgh, Pennsylvania 15260} 
\affiliation{Purdue University, West Lafayette, Indiana 47907} 
\affiliation{University of Rochester, Rochester, New York 14627} 
\affiliation{The Rockefeller University, New York, New York 10021} 

\affiliation{Istituto Nazionale di Fisica Nucleare, Sezione di Roma 1, $^v$Sapienza Universit\`{a} di Roma, I-00185 Roma, Italy} 

\affiliation{Rutgers University, Piscataway, New Jersey 08855} 
\affiliation{Texas A\&M University, College Station, Texas 77843} 
\affiliation{Istituto Nazionale di Fisica Nucleare Trieste/\ Udine, $^w$University of Trieste/\ Udine, Italy} 
\affiliation{University of Tsukuba, Tsukuba, Ibaraki 305, Japan} 
\affiliation{Tufts University, Medford, Massachusetts 02155} 
\affiliation{Waseda University, Tokyo 169, Japan} 
\affiliation{Wayne State University, Detroit, Michigan  48201} 
\affiliation{University of Wisconsin, Madison, Wisconsin 53706} 
\affiliation{Yale University, New Haven, Connecticut 06520} 
\author{T.~Aaltonen}
\affiliation{Division of High Energy Physics, Department of Physics, University of Helsinki and Helsinki Institute of Physics, FIN-00014, Helsinki, Finland}
\author{J.~Adelman}
\affiliation{Enrico Fermi Institute, University of Chicago, Chicago, Illinois 60637}
\author{T.~Akimoto}
\affiliation{University of Tsukuba, Tsukuba, Ibaraki 305, Japan}
\author{M.G.~Albrow}
\affiliation{Fermi National Accelerator Laboratory, Batavia, Illinois 60510}
\author{B.~\'{A}lvarez~Gonz\'{a}lez}
\affiliation{Instituto de Fisica de Cantabria, CSIC-University of Cantabria, 39005 Santander, Spain}
\author{S.~Amerio$^u$}
\affiliation{Istituto Nazionale di Fisica Nucleare, Sezione di Padova-Trento, $^u$University of Padova, I-35131 Padova, Italy} 

\author{D.~Amidei}
\affiliation{University of Michigan, Ann Arbor, Michigan 48109}
\author{A.~Anastassov}
\affiliation{Northwestern University, Evanston, Illinois  60208}
\author{A.~Annovi}
\affiliation{Laboratori Nazionali di Frascati, Istituto Nazionale di Fisica Nucleare, I-00044 Frascati, Italy}
\author{J.~Antos}
\affiliation{Comenius University, 842 48 Bratislava, Slovakia; Institute of Experimental Physics, 040 01 Kosice, Slovakia}
% ...

\collaboration{CDF Collaboration\footnote{With visitors from $^a$Universiteit Antwerpen, B-2610 Antwerp, Belgium, 
$^b$Chinese Academy of Sciences, Beijing 100864, China, 
$^c$University of Bristol, Bristol BS8 1TL, United Kingdom, 
$^d$University of California Irvine, Irvine, CA  92697, 
$^e$University of California Santa Cruz, Santa Cruz, CA  95064, 
$^f$Cornell University, Ithaca, NY  14853, 
$^g$University of Cyprus, Nicosia CY-1678, Cyprus, 
$^h$University College Dublin, Dublin 4, Ireland, 
$^i$University of Edinburgh, Edinburgh EH9 3JZ, United Kingdom, 
$^j$Universidad Iberoamericana, Mexico D.F., Mexico, 
$^k$University of Manchester, Manchester M13 9PL, England, 
$^l$Nagasaki Institute of Applied Science, Nagasaki, Japan, 
$^m$University de Oviedo, E-33007 Oviedo, Spain, 
$^n$Queen Mary, University of London, London, E1 4NS, England, 
$^o$Texas Tech University, Lubbock, TX  79409, 
$^p$IFIC(CSIC-Universitat de Valencia), 46071 Valencia, Spain,
$^x$Visitor from Royal Society of Edinburgh, Edinburgh, EH22PQ, United Kingdom.
%$^x$Royal Society of Edinburgh/Scottish Executive Support Research Fellow, 
}}
\noaffiliation
  % input author list
%\author{The CDF Collaboration}

\affiliation{URL: http://www-cdf.fnal.gov}

%\vspace*{1.5in}

\begin{abstract}
\noindent 
We present measurements of the forward-backward charge asymmetry in top pair 
production using $1.9 ~\ifb$ of $\ppbar$ collisions at $\sqrt{s}=1.96$ TeV 
recorded with the CDF II detector at the Fermilab Tevatron. Correcting for 
acceptance and measurement dilutions we obtain parton-level asymmetries of 
$\afbl = 0.17 \pm 0.08$ in the $\ppbar$ frame and $\afbt = 0.24 \pm 0.14$ 
in the $\ttbar$ frame. The values are consistent with the standard model 
expectation and disfavor exotic production mechanisms with significant 
negative values.
\end{abstract}

%activate the following line for publication
\pacs{ 12.38.Qk, 13.87.Ce, 14.65.Ha}

%for revtex
\maketitle

\date{\today}
The top quark, discovered in 1995 by both Tevatron experiments~\cite{topdis}, 
is the only known fermion with a mass of the order of the electroweak breaking 
scale. This suggests that it may play a special role in new physics. 
A detailed investigation of the production mechanism of top quarks will give 
insights into whether top quarks are produced via new physics processes.

In this Letter we present two analyses studying the forward-backward charge asymmetry 
of top quark pairs in $\ppbar$ collisions at $\sqrt{s}= 1.96$ TeV at the Fermilab Tevatron. 
In standard model QCD a charge asymmetry $\ac$ arises in next-to-leading order
$t\bar{t}X$ production. Because the strong interaction is invariant under
charge conjugation $\ac$ is equivalent to a forward-backward asymmetry $\afb$. 
%In standard model QCD arises in next-to-leading order $t\bar{t}X$ production 
%a charge asymmetry $\ac$, which is due to charge-conjugation
%invariance in strong interaction equivalent to a forward-backward asymmetry $\afb$. 
Recent calculations predict a slightly positive total $\afb = 5.0 \pm 1.5\%$ in 
the Tevatron $\ppbar$ rest frame \cite{kuhn2,afbother}, 
with the theoretical uncertainty driven by the size of corrections at higher orders. 
This small total $\afb$ combines a positive asymmetry from the interference of the 
Born and virtual (box) corrections ($\ttbar$) with a negative asymmetry from interference 
of initial and final state radiation amplitudes ($\ttbar+g$) \cite{afbtheory}. 

While the total $\afb$ value expected by the standard model is hardly measurable at the 
presently achievable precision, we are sensitive to large $\afb$ values 
(of order $\pm$30\%) predicted in some models with new physics, e.g. $Z'$-like states with 
parity violating couplings \cite{rosner} and theories with chiral color 
\cite{afbexotic,kuhn2}.
In contrast to searches for heavy resonances in the spectrum of the mass of
the top pair \cite{ttres}, a measurement of $\afb$ is sensitive to both
narrow and broad resonances. In addition, the presence of a massive
gluon may be visible in the asymmetry even above the collision energy due
to interference with the standard model gluon.  

Since a longitudinal boost changes the top quark direction, $\afb$ is frame dependent.
Undetected collinear gluon radiation makes the fundamental initial parton frame 
experimentally inacessible. However, the $\ttbar$ and the $\ppbar$ frame are
experimentally accessible and according to \cite{kuhn2} the $\afb$ values in the 
$\ppbar$ frame are predicted to be reduced by $\approx 30\%$ relative to the $\ttbar$ 
frame. 

We present here the first measurement of the top quark production $\afb$,
fully corrected to the parton level, in both the $\ppbar$ and $\ttbar$
frames. Correction to the intrinsic parton value allows direct comparison
to theoretical prediction, and measurements in two frames probe the
consistency and the frame dependence of the effect. A recent study~\cite{d0Afb}
measures a quantity which is related to the $\ttbar$ frame asymmetry but is
uncorrected for acceptance and resolution effects. The result $(12\pm 8\pm 1)\%$ 
is larger than expected, within errors, but difficult to interpret.

We use 1.9 $\ifb$ of $\ppbar$ collision data recorded by the CDF II detector. 
The detector is a forward-backward symmetric system consisting of a magnetic spectrometer 
surrounded by projective calorimeters and muon detectors \cite{CDF}. Charged 
track reconstruction in a 1.4 T axial field uses a large open cell drift chamber 
and silicon microstrip detectors for displaced secondary vertex detection. We 
use coordinates where $\phi$ is the azimuthal angle and $\theta$ is the polar 
angle with respect to the proton beam $z$-axis. Transverse energy is 
$E_T = E\sin\theta$, the rapidity is Y = $\frac{1}{2}\ln[(E+p_z)/(E-p_z)]$, and 
the pseudorapidity is $\rm{\eta = -ln[tan(\theta/2)]}$. 
  
We collect a sample of candidate events in the lepton+jets topology 
$\ttbar \rightarrow (W^{+}b)(W^{-}\bar{b})\rightarrow (q\bar{q}'b)
(\ell^-\bar{\nu}_{\ell}\bar{b})$ ~\cite{cc}, where one $W$-boson decays 
leptonically and the other hadronically, by triggering on a central 
($\abseta 1.0$) electron with $E_T>$18~GeV or central muon with transverse 
momentum $p_T>$18~GeV/$c$. After offline reconstruction we select events with 
an isolated electron with $E_T\geq 20$~GeV or muon with $p_T\geq 20$~GeV/$c$, 
missing transverse energy $\met \geq~$20~GeV ~\cite{metdef} consistent with a 
neutrino from $W$ decay, and at least four hadronic jets with $\abseta$~2.0 
and $\et\geq 20$~GeV. Jets are clustered in fixed cones of radius $\Delta R
\equiv\sqrt{(\Delta \eta)^2 + (\Delta \phi)^2} \leq 0.4$ and jet energies 
are corrected to parton-level values \cite{jes}. At least one jet must be 
$b$-tagged, i.e., contain a reconstructed secondary vertex consistent with the decay 
of a bottom hadron in the jet \cite{g4topxsec}. We find $484$ candidate events. 

The expected $\ttbar$ signal is studied using the {\sc pythia, herwig,} and {\sc mc@nlo} 
event generators~\cite{LOmc} and a full detector simulation \cite{cdfsim}. 
The top quark mass is set equal to $\mtop = 175 ~\gevcc$. 
The rates and kinematics of background processes are well modeled with simulation and data 
control samples \cite{topxsec} which will be discussed later. 
We expect a total of $87 \pm 23$ background events, leaving a $t\bar{t}$ signal 
of $397 \pm 32$ events, consistent with our previous cross-section measurement of 
$8.2\pm 1.0$ pb \cite{xsec82}.

\begin{table*}[t]
\caption{Measured asymmetries in large simulated $\ttbar$ samples.}
\begin{ruledtabular}
\begin{tabular}{lrrrr}
%\vspace*{-0.3cm}\\
%Generator           &\multicolumn{2}{c}{$\afbl $} &\multicolumn{2}{c}{$\afbt $}  \\
Generator           &       Parton-level $\afbl $   &     Reconstructed $\afbl $    &       Parton-level $\afbt $ &   Reconstructed $\afbt $  \\ \hline
{\sc Pythia}        &  $0.000 \pm 0.003  $       &  $-0.007 \pm 0.006 $       &     $ 0.000 \pm 0.001$   &   $-0.005 \pm 0.003$ \\
{\sc Herwig}        &  $0.000 \pm 0.006  $       &  $-0.013 \pm 0.012 $       &     $-0.003 \pm 0.002$   &   $-0.003 \pm 0.006$   \\
{\sc MC@NLO}        &  $0.038 \pm 0.002  $       &  $ 0.015 \pm 0.016$        &     $ 0.049 \pm 0.002 $   &   $0.017 \pm 0.007$    \\
\end{tabular}
\end{ruledtabular}
\label{t:simulations}
\end{table*}
Mass constraints on the $W$ bosons from top quark decay fix the jet parton assignment and
allow complete reconstruction of the $\ttbar$ kinematics. For the $\ppbar$ frame analysis we use 
the algorithm employed in the top quark mass measurement of Ref.~\cite{cdfreco}. 
Measured jet energies float within expected resolutions, $b$-tagged jets are taken as fragmented
$b$-quarks, 
both $W$ boson masses $M(\ell\nu)$ and $M(q\bar{q}')$ are constrained to $80.4 ~\gevcc$, 
and the top quark mass is constrained to $175 ~\gevcc$. For the $\ttbar$ frame analysis we 
use the technique described in Refs.~\cite{kareco} which employs constraints on
the $W$ boson masses, the reconstructed $t$-$\bar{t}$ mass difference (but not $\mtop$), the total 
transverse energy, and the $b$-likelihood of the jets \cite{jetprob}. In simulated $\ttbar$ 
samples the two procedures resolve the top direction with similar accuracy. The resolution on 
the direction of the hadronically decaying top quark $t_h$, expressed in terms of rapidity, is 
$\sigma_Y(t_h)\simeq 0.29$. The leptonically decaying top quark system $t_l$, which includes 
the indirectly measured neutrino, has $\sigma_Y(t_{\ell})\simeq 0.46$ and significant 
non-Gaussian tails (15\%). 

We measure the direction of the top quark in the $\ppbar$ center-of-mass frame using the 
cosine of the polar angle between the hadronic top quark and the proton beam, $\cos\alpha_p$.
The sign of the $t_h$ electric charge is opposite that of $Q_{\ell}$, the leptonic charge 
observed in the $t_l$ decay.
Assuming CP invariance, we find one top quark angle $\cos\theta = -Q_{\ell} \cdot \cos\alpha_p$ 
in each event and calculate the asymmetry in the $\ppbar$ center of mass frame \cite{tthesis}
\begin{equation}
\afbl = \frac{N(\cos\theta>0) - N(\cos\theta<0)}{N(\cos\theta>0) + N(\cos\theta<0)}.
\end{equation}                                 
This technique has the simplicity of relying only on the hadronic top quark reconstruction, 
but has the drawback of measuring asymmetries which are diluted by 30\% compared to the 
$\ttbar$ frame. 

The $\ttbar$ rest frame measurement exploits the Lorentz invariant difference between the 
$t$ and $\bar{t}$ rapidities $Y_t$ and $Y_ {\bar{t}}$. 
We use the reconstructed rapidity of $t_{\ell}$ and $t_h$ in each event, assume
CP invariance, and determine $\Delta Y = Y_t - Y_{\bar{t}} = Q_{\ell} \cdot(Y_{t_{\ell}}-Y_{t_h})$
from which we calculate the asymmetry in the approximate (LO) $\ttbar$ rest frame \cite{jthesis}
\begin{equation}
\afbt = \frac{N(\Delta Y > 0) - N(\Delta Y < 0)}{N(\Delta Y > 0) + N(\Delta Y < 0)}.   
\end{equation}
To connect this with other asymmetry measurements, we note that in 
the case of ideal resolution $\afbt$ reproduces the asymmetry measured in the equivalent 
Collins-Soper frame \cite{cs}. While it is sensitive to the larger $\ttbar$ frame asymmetry, 
$\Delta Y$ combines the uncertainties of both quark reconstructions, 
including the neutrino-related complications of the $t_{\ell}$ quark system. 

The expected measurement performance is evaluated using simulated samples. In 
Table~\ref{t:simulations} we compare asymmetries found after selection and 
reconstruction to parton-level asymetries calculated using perfect acceptance 
and resolution. The uncertainties reflect the simulation statistics. With the 
parton-shower generators {\sc pythia} and {\sc herwig} we see no intrinsic 
charge asymmetry at the parton-level, as expected, and verify that any 
forward-backward bias from selection and reconstruction is small. With the 
{\sc mc@nlo} generator, which includes the small QCD-induced charge asymmetry, 
we find parton-level values consistent with theoretical expectation in magnitude 
and the level of frame dependence. With large statistics the measured values are 
sensitive to the small asymmetry, but diluted by acceptance and reconstruction effects. 
Dilution corrections, as well as the expected sensitivity in our finite dataset, 
are discussed later. 
The calibration of the simulation to the physical detector geometry and acceptance 
has been checked in studies of electroweak processes \cite{EWchecks}. For example, 
the leptonic charge asymmetry in $W^{\pm}\rightarrow l^{\pm}\nu$ agrees with our 
simulated physics and detector model within the statistical uncertainty of 
$\approx 0.004$. 

The $\afb$ measured in data must be corrected for background contributions which 
include asymmetric weak processes.
$W$ + jets events with tagged heavy flavor ($W$+hf) or mistagged light partons 
($W$+lf) are modeled using {\sc alpgen} \cite{alpgen} interfaced to {\sc pythia} 
parton showering, along with $b$-tagging and mistagging rates pa\-ra\-me\-terized from 
jet data. Small electroweak backgrounds (EW), $WW, WZ, ZZ$ and single-top, are 
modeled with {\sc pythia} and with {\sc madevent} \cite{mad} respectively.
%are modeled with {\sc pythia} and single-top with {\sc madevent} \cite{mad}. 
The non-$W$ (QCD) electron background is studied using data events with five 
jets where one jet models a misreconstructed electron; the same sample is used 
for non-$W$ muons after re-weighting the lepton acceptance. The background levels 
and asymmetries expected in the two analyses are shown in Table ~\ref{t:backgrounds}. 
The combined results are listed in the last row. 
\begin{table}[h]
\caption{Backgrounds, estimated number of events $N_{exp}$ and their effective asymmetries.}
\begin{ruledtabular}\begin{tabular}{llrr}
 Process  & $N_{exp}$ & $\afbl $ &  $\afbt $\\
\vspace*{-0.3cm}\\
$W$+hf &  $37 \pm 10   $ & $ -0.087 \pm 0.005  $ &  $-0.045 \pm 0.003 $\\
$W$+lf &  $20 \pm 5    $ & $  0.044 \pm 0.008   $ & $-0.006 \pm 0.015  $\\
EW     &  $12 \pm 1      $ & $ -0.022 \pm 0.014   $ & $-0.015 \pm 0.044  $\\
QCD    &  $18 \pm 16   $ & $ -0.008 \pm 0.004   $ & $ 0.006 \pm 0.010  $\\
\hline
Total  &  $87 \pm 23   $ & $ -0.053 \pm 0.004   $ & $-0.021 \pm 0.007  $\\
\end{tabular}
\end{ruledtabular}
\label{t:backgrounds} \rm
\end{table}
\begin{figure*}[ht]
\begin{center}
\begin{minipage}[t]{5.5in}
\epsfxsize=2.5in
\epsffile{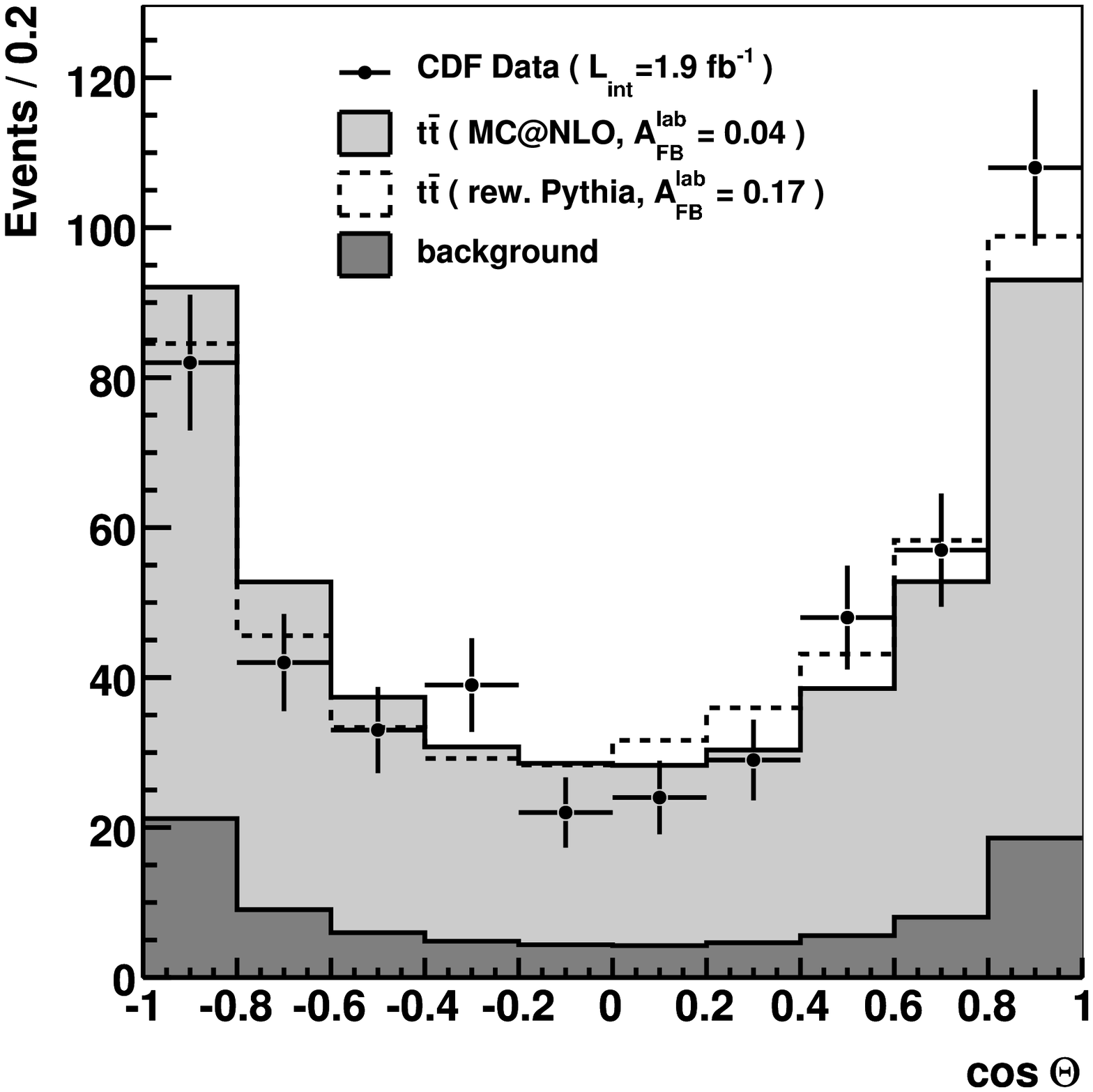}
\hfill
\epsfxsize=2.5in
\epsffile{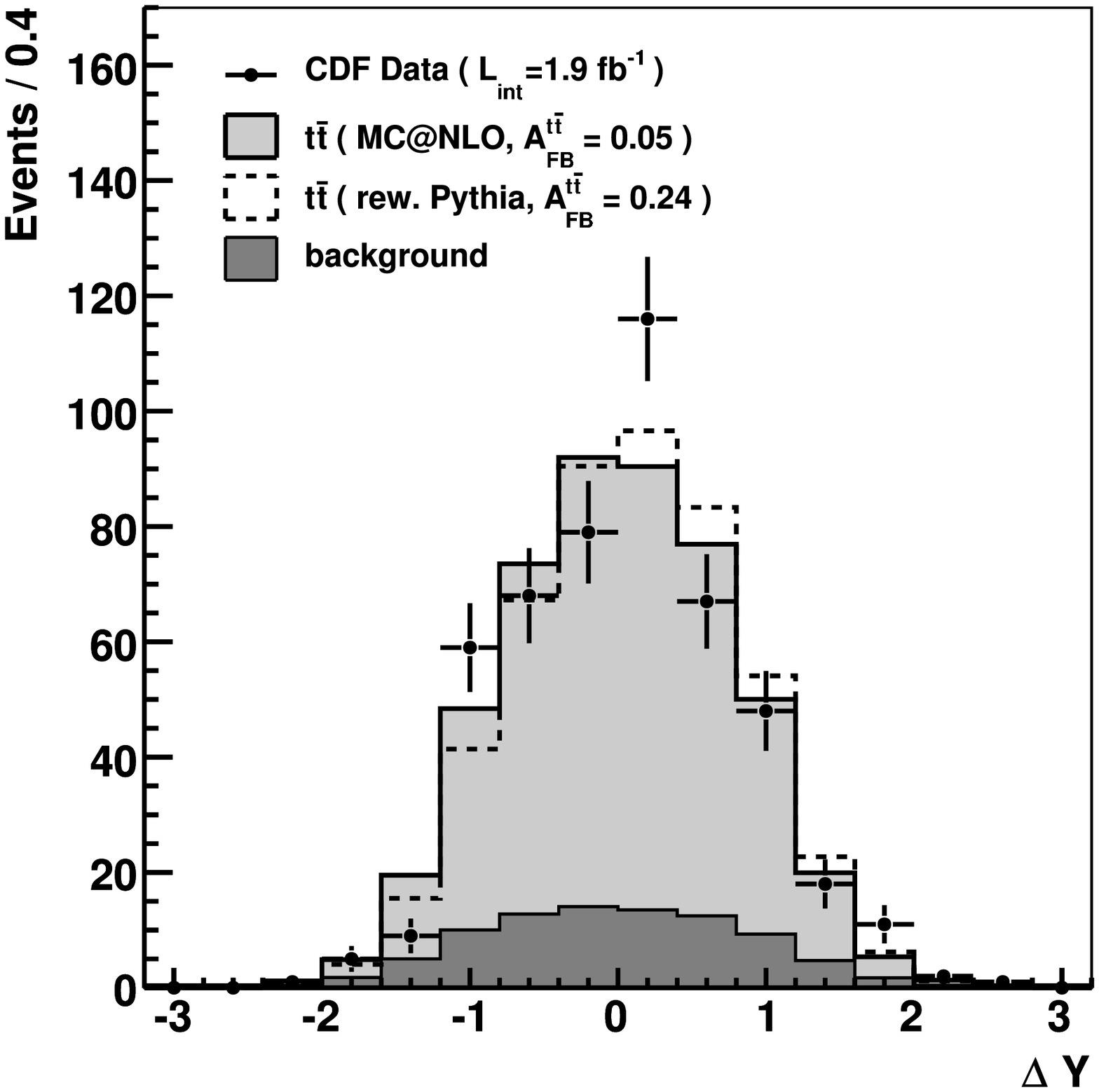}\\
\caption{\label{f:reco} {\small The two top quark production angle variables, $\cos\theta$ 
for the $\ppbar$ frame and ~$\Delta Y$ for the $\ttbar$ frame. The solid line is the prediction for
$\ttbar$ with {\sc mc@nlo} model of the QCD induced charge asymmetry and $\sigma_{\ttbar}=8.2$ pb,
plus the expected non-$\ttbar$ backgrounds. The dashed curve shows the prediction when $\ttbar$ 
is reweighted according to the form $1 + \afb\cos\alpha$ using measured values of $\afb$.}}
\end{minipage}
\end{center}
\end{figure*}          
\begin{table*}[ht]
\caption{Predicted ({\sc mc@nlo} + non-$\ttbar$) and measured asymmetries without further
corrections.}
\begin{ruledtabular}\begin{tabular}{lcrlrr}
%                    & \multicolumn{2}{c}{$\afbl $}    & \multicolumn{2}{c}{$\afbt $ } \\
                    & $N_{jets}$ &       Predicted $\afbl $    &    Measured $\afbl $    &  Predicted $\afbt $  & Measured $\afbt $ \\ \hline
%Reconstructed Data &&&&\\
Recontructed Data & $\ge 4$ &  $ 0.001 \pm 0.010$          &   $0.099\pm 0.045$  &  $0.010\pm 0.007$   &  $0.087\pm 0.045$ \\[0.2cm]
%Background Subtracted &&&&\\
Background Subtracted & $ \ge 4$ &  $0.015\pm 0.016$     &  $0.130\pm 0.055$  &  $0.017\pm 0.007$     &    $0.119\pm 0.064$ \\
& $  = 4$ &  $0.032\pm 0.018$      &  $0.120\pm 0.064$  &  $0.038\pm 0.008$     &    $0.132\pm 0.075$ \\
& $ \ge 5$ & $-0.027\pm 0.032$     &  $0.160\pm 0.109$  &  $ -0.033\pm 0.012$  &    $0.079\pm 0.123$  \\
\end{tabular}
\end{ruledtabular}
\label{t:raw_data}
\end{table*}

Fig.~\ref{f:reco} shows the measured distributions of $\cos\theta$ and $\Delta Y$ in the 484 
$b$-tagged $\ttbar$ candidates, along with predictions based on simulated $\ttbar$ events 
from the {\sc mc@nlo} generator in combination with our non-$\ttbar$ background models. 
The measured asymmetries are displayed in Table~\ref{t:raw_data}. The background-corrected 
values, derived by subtracting the composite model shape bin-by-bin, show a positive 
asymmetry which is larger than but consistent with the {\sc mc@nlo} predictions within 
uncertainties. Our background-corrected $\afbt$, although measured in a slightly different
visible phase space, is very consistent with the measurement from the D\O\ 
collaboration~\cite{d0Afb}.
Subdividing the data by lepton types and lepton charges shows a consistent positive 
asymmetry across all samples.

To study the two contributions $t\bar{t}$ and $t\bar{t}+g$ with different expected 
sign in $\afb$, we split our data sample into events without any additional hard jet 
($N_{jets}=4$, 85\% $\ttbar$) and events with at least one additional hard jet 
($N_{jets}\ge 5$, 53\% $\ttbar + g$). Our background-corrected $\afbl$ and $\afbt$
values for this study are presented in Table~\ref{t:raw_data}. The $N_{jets}$
dependence is not as strong as seen in ~\cite{d0Afb}, but the limited statistics
does not allow a firm conclusion. 

The distributions in Fig.~\ref{f:reco} are distorted from their true parton-level
shapes by acceptance bias and reconstruction errors. We use a 
matrix inversion technique to derive the parton-level distributions and $\ttbar$ 
asymmetries. If an event in bin $j$ at parton-level is collected 
with efficiency $\epsilon_j$ and migrates to bin $i$ at the measurement level 
with probability $S_{ij}$, the bin-by-bin parton-level distributions $P_j$ can 
be found from the background-corrected data distributions $D_{i}$ by the 
inverse transformation
\begin{equation}
P_j =  \epsilon_j^{-1}\cdot S_{ji}^{-1}\cdot D_{i}.
\end{equation}
We simplify each distribution to four bins, 
with two bins on either side of the cross-over at $\cos\theta=\Delta Y = 0$. The efficiencies and migration matrix $S_{ij}$ are 
derived by comparing the parton and reconstructed level quantities using the zero asymmetry {\sc pythia} 
$\ttbar$ simulations. In the $\cos\theta ~(\Delta Y)$ analysis roughly $13\% ~(25\%)$ of events 
change signs, but the matrix is symmetric within uncertainties. The symmetry of the matrix, which 
follows from the forward-backward symmetry of the detector, ensures that the
inversion is insensitive to small errors in the modeling of the migration parameters.

The expected performance of the complete calculation is evaluated with simulated samples.
Sensitivity to the asymmetry model is studied using {\sc pythia} samples 
that have been reweighted in the top quark production angle for a range of possible asymmetry 
functions and magnitudes varying between 0.0 and 0.30. Sensitivity to the QCD-induced asymmetry 
is studied with {\sc mc@nlo}. The effect of extra jets is studied with exclusive 
$\ttbar + 0$ ~parton and $\ttbar + 1$ ~parton samples made with the {\sc alpgen} generator. 
Each sample was reconstructed, measured, and propagated back to the parton 
level with the procedures described above. 
For all conditions the procedure 
returns mean values within 0.02 of the true value. The predicted  
statistical precisions in our 1.9 $\ifb$ dataset are $\delta \afbl = 0.09$ and 
$\delta \afbt = 0.13$. 

Additional sources of uncertainty are evaluated using simulated samples with reasonable 
variations on the assumptions for background shape and normalization, signal shapes, 
the top quark mass, the parton distribution functions, the amount of initial and final 
state gluon radiation, and the calorimeter energy scale. 
The largest uncertainty in the $\afbl$ analysis is the background normalization and the 
largest in the $\afbt$ analysis is the $\Delta Y$ shape modeling, being roughly 
$\delta\afb \simeq 0.02$ in each. The total systematic uncertainty is $\delta\afb = 0.04$ 
for both techniques.
 
Applying our algorithm to the inclusive background-subtracted distributions in Fig.~\ref{f:reco}, 
we find parton level asymmetries of $\afbl = 0.17 \pm 0.07 \pm 0.04$ and $\afbt = 0.24 
\pm 0.13 \pm 0.04$, where the uncertainties are statistical and systematic, respectively. 
In Fig.~\ref{f:reco}, the dashed lines show that the data are in good agreement with models 
derived by reweighting the generated top quark production angle $\alpha$ in the symmetric 
{\sc pythia} sample with form $1 + \afb\cos\alpha$ using the measured $\afb$. 

In conclusion, we have measured a forward-backward and (equivalent) charge asymmetry in a strong 
process at high energy using reconstructed $\ttbar$ events produced in $\ppbar$ collisions 
at $\sqrt{s}$ = 1.96 TeV. We find forward-backward parton-level asymmetries of 
$\afbl = 0.17 \pm 0.08$ in the $\ppbar$ frame and $\afbt = 0.24 \pm 0.14$ in the $\ttbar$ frame.
Our results show the expected frame dependence, are consistent ($\leq 2\sigma$) 
with the small ($\sim 0.05$) charge asymmetry expected from QCD, and they disfavor 
exotic sources of top quark production with significant negative $\afb$ values~\cite{kuhn2}.

We thank the Fermilab staff and the technical staffs of the participating institutions for their vital contributions. This work was supported by the U.S. Department of Energy and National Science Foundation; the Italian Istituto Nazionale di Fisica Nucleare; the Ministry of Education, Culture, Sports, Science andTechnology of Japan; the Natural Sciences and Engineering Research Council of Canada; the National Science Council of the Republic of China; the Swiss National Science Foundation; the A.P. Sloan Foundation; the Bundesministerium f\"ur Bildung und Forschung, Germany; the Korean Science and Engineering Foundation and the Korean Research Foundation; the Science and Technology Facilities Council and the Royal Society, UK; the Institut National de Physique Nucleaire et Physique des Particules/CNRS; the Russian Foundation for Basic Research; the Ministerio de Educaci\'{o}n y Ciencia and Programa Consolider-Ingenio 2010, Spain; the European Community's Human Potential Programme; the Slovak R\&D Agency; and the Academy of Finland.

%\vspace*{7.0in}

%\bibliography{afb_prl}

\begin{thebibliography}{99}
\bibitem{cc} The inclusion of the charge conjugate process should be assumed.
\bibitem{metdef} Missing transverse energy, $\met$, is defined 
as the magnitude of the vector $ -\sum_i E_T^i \vec n_i$ where $E_T^i$ are the magnitudes of transverse energy contained in each calorimeter tower $i$, and $\vec n_i$ is the unit vector from the interaction vertex to the tower in the transverse ($x,y$) plane.
\end{thebibliography}

% Phys.~Rev.~Lett. {\bf },
% Phys. Rev. {\bf D},

\end{document}